\documentclass{egpubl}
\usepackage{eurovis2026}

\SpecialIssuePaper         %

\CGFccby

\usepackage[T1]{fontenc}
\usepackage{dfadobe}  
\usepackage{soul}
\usepackage{enumitem}
\usepackage{stfloats}
\usepackage{xcolor}
\usepackage{booktabs}
\usepackage{setspace}

\usepackage{cite}  %

\BibtexOrBiblatex
\electronicVersion
\PrintedOrElectronic
\usepackage{graphicx}
\ifpdf\pdfcompresslevel=9\fi

\usepackage{egweblnk}
\usepackage{xcolor}
\usepackage{tabularx}
\usepackage{booktabs}
\usepackage{setspace}
\usepackage[table]{xcolor}
\usepackage[most]{tcolorbox} 
\usepackage{hyperref}
\usepackage{makecell}

\usepackage{subcaption}
\usepackage{graphicx}

\definecolor{linkColor}{HTML}{257E98}
\definecolor{verylightblue}{RGB}{220,240,255}
\newtcolorbox{takeawaybox}{
  breakable,
  colback=verylightblue,
  colframe=linkColor!50!black,
  boxrule=0.4pt,
  arc=2mm,
  left=8pt,right=8pt,top=6pt,bottom=6pt
}

\newtcbox{\takeaway}{
  on line,
  colback=verylightblue,
  colframe=linkColor!50!black,
  boxrule=0.4pt,
  arc=2pt,
  left=3pt,right=3pt,top=1.5pt,bottom=1.5pt
}

\newcommand{\secref}[1]{\hyperref[#1]{Sec.~\ref*{#1}}}
\newcommand{\appendixref}[1]{\hyperref[#1]{Appendix~\ref*{#1}}}
\newcommand{\figref}[1]{\hyperref[#1]{Fig.~\ref*{#1}}}
\newcommand{\eqnref}[1]{\hyperref[#1]{Eqn.~\ref*{#1}}}
\newcommand{\tabref}[1]{\hyperref[#1]{Table ~\ref*{#1}}}

\definecolor{quoteColor}{HTML}{ff5733}

\def\subsubsec#1
{\subsubsection{#1}}

\setuldepth{Berlin}
\definecolor{linkColor}{HTML}{257E98}

\definecolor{verylightblue}{RGB}{220,240,255} %
\definecolor{verylightgreen}{RGB}{220,252,225} %

\newcommand{\customQuote}[2]{\sethlcolor{#1}\hl{``#2''}}

\newcommand{\quoteP}[1]{\customQuote{verylightblue}{#1}}

\newcommand{\quoteE}[1]{\customQuote{verylightgreen}{#1}}

\newcommand{\customHighlight}[2]{%
  \sethlcolor{#1}%
  \hl{#2}%
}
\newcommand{\hlP}[1]{%
  \customHighlight{verylightblue}{#1}%
}

\newcommand{\hlE}[1]{%
  \customHighlight{verylightgreen}{#1}%
}

\definecolor{tblheader}{gray}{0.93}
\definecolor{tblstripe}{gray}{0.97}
\newcolumntype{Y}{>{\raggedright\arraybackslash}X}

\usepackage{tikz,xcolor}
\definecolor{circnumlight}{RGB}{235,244,255} %
\newcommand{\circnum}[2][1.4em]{%
  \tikz[baseline=(num.base)]%
    \node[draw, circle, fill=circnumlight, minimum size=#1,
      inner sep=0pt, line width=0.6pt] (num) {\sffamily\bfseries\fontsize{11pt}{11pt}\selectfont #2};}

\definecolor{takeawayBg}{RGB}{248,248,246}       %
\definecolor{takeawayBorder}{RGB}{230,232,235}   %
\colorlet{takeawayLabel}{linkColor!60!black}     %

\newcommand{\tkLabel}{%
  \textcolor{takeawayLabel}{\underline{\textsc{\fontseries{sb}\selectfont Takeaway.}}}\hspace{1pt}%
}

\newcommand{\Takeaway}[1]{%
  \par\noindent
  \begingroup
  \setlength{\fboxsep}{1.1pt}%
  \setlength{\fboxrule}{0.25pt}%
  \fcolorbox{takeawayBorder}{takeawayBg}{%
    \hspace*{-\fboxsep}%
    \parbox{\linewidth}{%
      \noindent\tkLabel~#1%
    }%
    \hspace*{-\fboxsep}%
  }%
  \endgroup
  \par
}

\newcommand{\circnumtable}[2][1.12em]{%
  \tikz[baseline=(num.base)]%
    \node[
      draw,
      circle,
      fill=circnumlight,
      minimum size=#1,
      inner sep=0pt,
      line width=0.5pt
    ] (num) {\sffamily\bfseries\fontsize{8.5pt}{8.5pt}\selectfont #2};%
}

\colorlet{supMarkFill}{linkColor!8}
\colorlet{supMarkStroke}{linkColor!45!black}
\colorlet{supMarkText}{linkColor!80!black}

\let\oldthebibliography\thebibliography
\renewcommand\thebibliography[1]{%
  \oldthebibliography{#1}%
  \setlength{\parskip}{0pt}%
  \setlength{\itemsep}{0.5pt}%
  \setlength{\parsep}{0pt}%
}

\title[Designing Annotations in Visualization]%
      {Designing Annotations in Visualization: \\Considerations from Visualization Practitioners and Educators}

\author[Md Dilshadur Rahman \& Devin Lange \& Ghulam Jilani Quadri \& Paul Rosen]
{\parbox{\textwidth}{\centering
Md\,Dilshadur Rahman$^{1}$\orcid{0009-0008-5467-615X} \quad
Devin Lange$^{2}$\orcid{0000-0002-3467-0294} \quad
Ghulam\,Jilani Quadri$^{3}$\orcid{0000-0002-8054-5048} \quad
Paul Rosen$^{1}$\orcid{0000-0002-0873-9518}
}\\
\parbox{\textwidth}{\centering
$^1$Scientific Computing and Imaging Institute, University of Utah, USA\\
$^2$Harvard Medical School, USA\\
$^3$University of Oklahoma, USA
}
}

\begin{document}

\teaser{
 \includegraphics[width=1\linewidth]{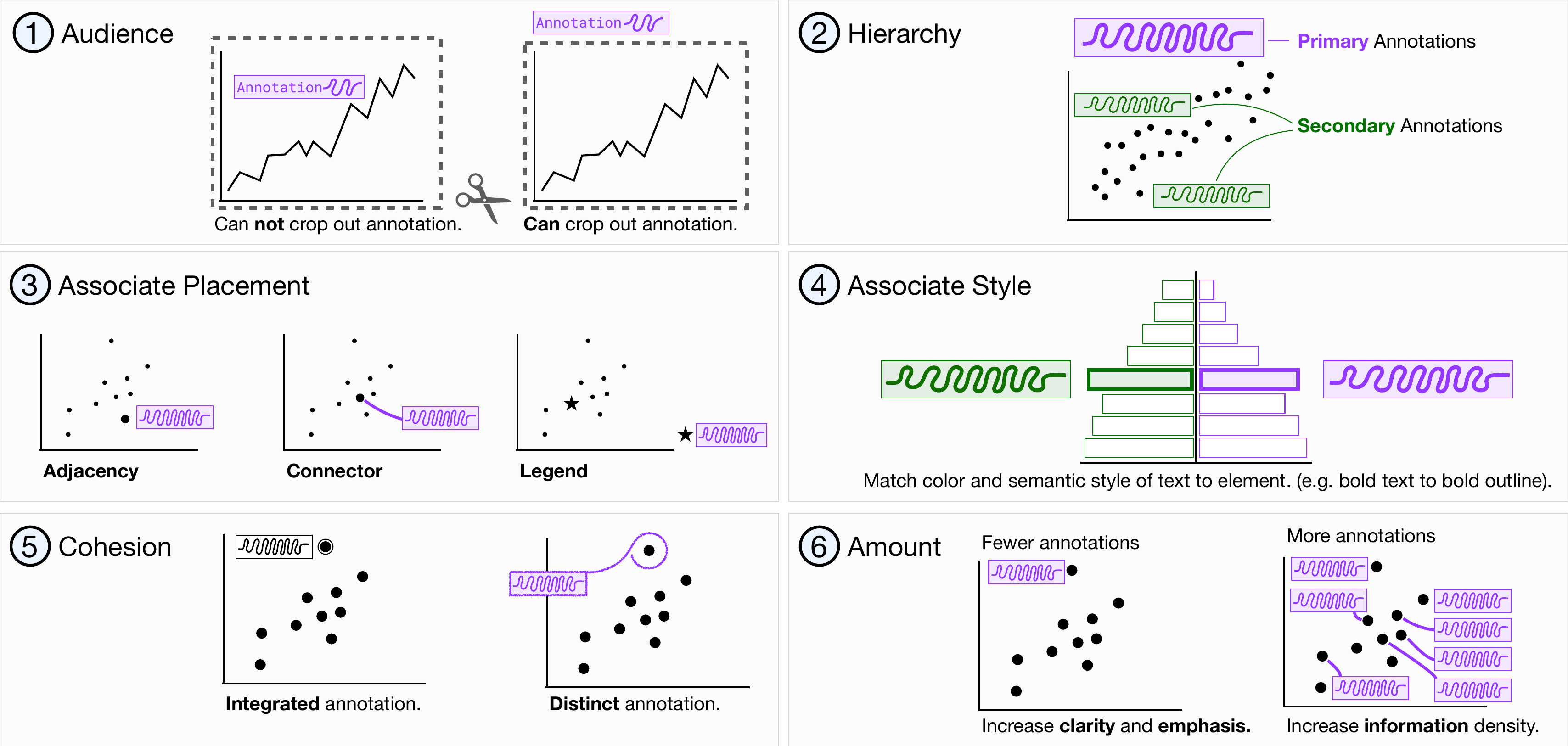}
 \centering
\vspace{1pt}
\caption{The sketches illustrate six annotation design considerations derived from interviews with practitioners (P1--P10) and educators (E1--E7). \protect\circnumtable[1.05em]{1} \textbf{Audience:} decide what context must appear on the chart so it remains interpretable when reused without surrounding text. \protect\circnumtable[1.05em]{2}~\textbf{Hierarchy:} make the primary annotation dominant and secondary notes visually subordinate. \protect\circnumtable[1.05em]{3} \textbf{Placement:} place text next to its target when possible, use short connectors when proximity is infeasible, and use a key or legend only when direct attachment would clutter the view. \protect\circnumtable[1.05em]{4} \textbf{Association:} use consistent cues, such as color and styling, to link annotation text to its target data element. \protect\circnumtable[1.05em]{5} \textbf{Cohesion:} decide whether annotations blend with the chart or stand apart as a distinct layer. \protect\circnumtable[1.05em]{6} \textbf{Amount:} keep only the annotations necessary to support the main message, removing those that clutter the view. We provide a companion gallery that maps public professional examples from a published dataset~\cite{rahman2023exploring} to these considerations: \href{https://vis-annotations.github.io/annotation-design/}{vis-annotations.github.io/annotation-design}.}
\label{fig:teaser}
\vspace{2pt}
}

\maketitle
\begin{abstract}
   Annotation is a central mechanism in visualization design that enables people to communicate key insights. Prior research has provided essential accounts of the visual forms annotations take, but less attention has been paid to the decisions behind them. This paper examines how annotations are designed in practice and how educators reflect on those practices. We conducted a two-phase qualitative study: interviews with ten practitioners from diverse backgrounds revealed the heuristics they draw on when creating annotations, and interviews with seven visualization educators offered complementary perspectives situated within broader concerns of clarity, guidance, and viewer agency. These studies provide a systematic account of annotation design knowledge in professional settings, highlighting the considerations, trade-offs, and contextual judgments that shape the use of annotations. By making this tacit expertise explicit, our work complements prior form-focused studies, strengthens understanding of annotation as a design activity, and points to opportunities for improved tool and guideline support.

\printccsdesc   
\end{abstract}  
\section{Introduction}
\label{sec.intro}

Amanda Cox, former Graphics Editor of The New York Times and a leading figure in data journalism, once remarked, \textit{``the annotation layer is the most important thing we do… otherwise it's a case of here it is, you go figure it out''} \cite{kirk2012data, ren2017chartaccent}, underscoring that annotations are central to how visualizations convey meaning. We use \emph{annotations} to refer to textual or graphical elements that form an author-supplied communicative layer over the underlying encodings, directing attention, highlighting salient features, providing context, and guiding interpretation~\cite{rahman2024qualitative,munzner2014visualization}. Annotations can improve comprehension, recall, and engagement~\cite{borkin2013makes,ajani2021declutter,zhi2019linking,bancilhon2023combining,6327259,bryan2020analyzing}, support mental model construction of data~\cite{cedilnik2000procedural}, and externalize analytic findings for collaboration and narrative communication~\cite{kang2014characterizing,sevastjanova2021visinreport,romat2019activeink,chen2011supporting,mahyar2014supporting,segel2010narrative,hullman2011visualization,satyanarayan2014authoring,ge2020canis}.

Visualization guidance discusses annotation-related design choices such as labeling, emphasis, and narrative framing~\cite{munzner2014visualization,Ware:2004:IVP,tufte2001visual,cairo2016truthful,strobelt2015guidelines,segel2010narrative}, and research has examined annotations via professional visualization analyses~\cite{rahman2023exploring,segel2010narrative}, authoring tools and grammars~\cite{ren2017chartaccent,hullman2013contextifier,rahman2025annogram,chen2025chartmark,latif2019authoring,spritzer2015towards}, artifact analyses~\cite{rahman2024qualitative,stokes2025analysis}, and audience effects in chart and text integration~\cite{stokes2022striking,borkin2015beyond}. These efforts clarify common forms of annotation, document audience effects in specific settings, and propose systems that scaffold annotation authoring. However, they provide limited empirical visibility into how visualization authors make annotation decisions in context. In practice, visualization authors must decide what evidence warrants an annotation, what context or caveats must be made explicit in the visualization itself, and how to place and style annotations while managing attention and annotation density. These choices are consequential: annotations are often the primary channel through which authors communicate intended interpretation alongside the visual encodings. Without an account of this decision logic, annotation guidance can remain abstract: visualization authors may learn principles, but still lack support for the decisions and trade-offs required to produce effective annotations within the constraints of a given visualization, audience, tools, and medium.

We use the term \emph{practitioner} to denote professionals who regularly design, produce, or communicate with data visualizations, such as data journalists, analysts, visualization designers, and consultants. Their annotation choices shape how audiences interpret charts, yet the reasoning behind these choices remains largely undocumented. Design scholarship argues that expert work relies on tacit, situated knowledge, and that making such knowledge explicit can surface the considerations and trade-offs that organize practice and support practice-grounded tool and pedagogy design~\cite{polanyi2009tacit,schon2017reflective,suchman1987plans}. Prior practice-grounded studies have characterized visualization design workflows and judgment in professional settings~\cite{parsons2021understanding,parsons2020design,parsons2020data,parsons2021fixation,alspaugh2018futzing,bigelow2016iterating,bigelow2014reflections,hoffswell2020techniques,baigelenov2025visualization,parsons2025judgment}, but comparable accounts of annotation-specific decision-making remain underexplored.

To address this gap, we conducted a two-phase study that brings practitioners' accounts of annotation design into dialogue with educators' reflections on those accounts. We include educators because they regularly translate annotation guidance into critique and instruction, and can help surface boundary conditions and risks that practitioners may not articulate explicitly. In \textbf{Phase 1}, we interviewed ten \hlP{practitioners} from diverse domains who regularly create annotated visualizations and elicited the considerations that guide their everyday annotation decisions. In \textbf{Phase 2}, we interviewed seven \hlE{visualization educators and researchers}, presented the practitioner-derived considerations, and asked them to reflect on boundary conditions, risks, and teaching-oriented framings. The aim was not to validate practitioners or privilege one group over the other, but to use structured peer debriefing to surface key strategies, tensions, and trade-offs in annotation design~\cite{spall1998peer}.

We make three contributions. \textbf{First,} we provide a practice-grounded account of annotation design in professional visualization work, distilled into six design considerations that capture recurring annotation-specific decision points and trade-offs with examples at \href{https://vis-annotations.github.io/annotation-design/}{vis-annotations.github.io/annotation-design}. \textbf{Second,} we juxtapose practitioner and educator perspectives, showing where they align, where they diverge, and which boundary conditions and tensions these differences reveal for annotation practice. \textbf{Third,} we outline implications for visualization design, tool support, and teaching, treating these considerations as a shared vocabulary for design critique, tool development, and future empirical work on annotated visualizations. Throughout, we present these considerations as heuristics for reflection rather than universal prescriptions.

\section{Related Work}
\label{sec.related-work}
\paragraph*{Annotations in Visualizations.}
Annotations augment a visualization with author-supplied text and graphics that reference marks, regions, or coordinate space and shape how viewers connect evidence to interpretation~\cite{munzner2014visualization,rahman2024qualitative}. Prior work has formalized annotation design spaces by characterizing forms, functions, and referential scope, including distinctions between annotations that introduce external context and those that emphasize patterns already present in the data~\cite{hullman2013contextifier,ren2017chartaccent,rahman2024survey}. Complementary frameworks relate annotation use to task structure in visualization activity~\cite{Brehmer2013,rahman2024exploring} and conceptualize annotations as overlays and cues that support communication when their role is visually clear and consistently applied~\cite{kong2012graphical,kong2017internal}.

Empirical studies show that annotation choices affect reader outcomes and hinge on specific design parameters. Chart--text integration work demonstrates that the amount and placement of text relative to marks can shift what readers take away from a visualization~\cite{kim2021towards,stokes2022striking,zhi2019linking}. Research on emphasis and overlays shows benefits when highlights are selective and role-consistent, alongside risks of distraction and cue interference when salience is overused~\cite{kong2009perceptual,kong2012graphical,strobelt2015guidelines}. Narrative visualization research further demonstrates how annotation and accompanying text structure reading order and frame interpretation in explanatory settings~\cite{segel2010narrative,hullman2011visualization,satyanarayan2014authoring,ge2020canis,lee2015more}. Adjacent work highlights constraints that shape annotation design in practice, including accessibility and legibility~\cite{elavsky2022accessible,joyner2022visualization} and risks of misleading communication in public-facing contexts~\cite{fan2022annotating,lisnic2023misleading,correll2019ethical}.

Systems and grammars operationalize parts of this space by treating annotations as explicit authoring constructs with targets and layout behavior~\cite{ren2017chartaccent,hullman2013contextifier,rahman2025annogram,chen2025chartmark}. Other work foregrounds production constraints that shape annotation authoring, including limited space in dashboards and adaptation across media and screen sizes~\cite{elias2012annotating,badam2022integrating,hoffswell2020techniques,kim2022cicero}. Yet this body of work offers limited empirical visibility into how designers coordinate placement, density, emphasis, and framing across revision cycles constrained by reuse and tooling. Our study addresses this gap by centering practitioners' accounts of how these factors are weighed in everyday work, and by incorporating educators' boundary-condition critiques to articulate design considerations that remain underexamined in prior research.

\paragraph*{Tacit Knowledge and Visualization Design.}
Design and human--computer interaction scholarship characterizes expert knowledge as tacit and situated, enacted through iterative judgment under material and organizational constraints, in which designers adapt through feedback and negotiate trade-offs rather than follow fixed prescriptions~\cite{polanyi2009tacit,schon2017reflective,cross1982designerly}.
Within the visualization community, scholarship has traditionally emphasized visible artifacts such as marks, encodings, and interactions, with comparatively less attention to the reasoning that produces them. Practice-grounded studies of visualization design cognition show how practitioners draw on precedent, reframe problems, and exercise judgment while negotiating interpretability, aesthetics, and stakeholder needs~\cite{parsons2021understanding,parsons2020design,parsons2021fixation,baigelenov2025visualization,parsons2025beyond}. Design study methodology offers an established route for surfacing design reasoning through work in context with explicit process reporting and rigor criteria~\cite{sedlmair2012design,meyer2019criteria}. Prior accounts rarely treat annotation decisions as a distinct form of design reasoning or examine the judgments that produce them. Our study adopts the tacit/situated lens to motivate eliciting practitioners' judgments directly, using educator reflection as structured peer debriefing to articulate the boundary conditions, risks, and trade-offs that practitioners navigate but seldom make explicit~\cite{spall1998peer}.

\section{Methodology}
\label{sec.method}
\definecolor{tblheader}{gray}{0.93}
\definecolor{tblstripe}{gray}{0.97}

\begin{table*}[t]
\caption{Practitioner participants (P1--P10).}
\label{tab:participants}
\centering

\scriptsize
\setlength{\tabcolsep}{2.2pt} %
\renewcommand{\arraystretch}{1.02}
\rowcolors{2}{gray!6}{white}

\resizebox{\textwidth}{!}{%
\begin{tabular}{l l c l l c l l l}
\toprule
\rowcolor{gray!12}
\textbf{\textsc{ID}} &
\textbf{\textsc{Role}} &
\textbf{\textsc{Exp.}} &
\textbf{\textsc{Gender}} &
\textbf{\textsc{Region}} &
\textbf{\textsc{Vis edu.}} &
\textbf{\textsc{Vis types}} &
\textbf{\textsc{Tools}} &
\textbf{\textsc{Audience}} \\
\midrule
P1  & Data analyst             & 4--6  & Man   & Africa        & Y & Dash,Int,Story,Rep           & BI,TB,XL,PPT     & External clients; students/training \\
P2  & Graphics editor          & 4--6  & Woman & Europe        & Y & Int,Story,Oth                & DW,AI,AE,QG      & General public \\
P3  & Visualization manager    & 7--10 & Man   & North America & Y & Dash,Int,Sci,Pub,Story,Rep   & PY,DW,IG,AI,D3   & Government/policy; general public \\
P4  & Visualization consultant & 10+   & Man   & Europe        & Y & Dash,Int,Sci,Story           & BI,FL,D3,VG      & External clients; general public \\
P5  & Project manager          & 1--3  & Woman & North America & Y & Pub,Story,Rep                & FL,FG            & Government/policy \\
P6  & Data analyst             & 7--10 & Woman & North America & Y & Dash,Int,Rep                 & BI,TB            & Within-organization stakeholders \\
P7  & Data director            & 4--6  & Woman & North America & Y & Story,Rep                    & R,GS             & Government/policy; general public \\
P8  & University professor     & 10+   & Man   & North America & N & Sci,Story,Oth                & R,PPT            & Technical audiences; students/training \\
P9  & Visualization designer   & 4--6  & Woman & Asia          & Y & Dash,Int,Pub,Story           & FL,FG,AI         & General readers; technical audiences \\
P10 & Analytics lead           & 10+   & Other & North America & Y & Dash,Int,Pub,Story,Rep       & R,D3,VL,OB,SB    & Within-organization stakeholders \\
\bottomrule
\end{tabular}%
}

\vspace{2pt}
\begin{minipage}{\textwidth}
\scriptsize\raggedright
\textit{\textbf{Abbrev.}}~~
\textbf{\textsc{Exp.}}~years working with visualizations;~~
\textbf{\textsc{Vis edu.}}~formal visualization education (Y/N).\\[3pt]
\textbf{\textsc{Vis types:}}~~
Dash=dashboards;~
Int=interactive;~
Sci=scientific/technical;~
Pub=public communication;~
Story=data storytelling;~
Rep=stakeholder reports;~
Oth=other.\\[3pt]
\textbf{\textsc{Tools:}}~~
BI=Power BI;~
TB=Tableau;~
XL=Excel;~
PY=Python;~
R=R;~
DW=Datawrapper;~
IG=Infogram;~
AI=Illustrator;~
AE=After Effects;~
QG=QGIS;~
FL=Flourish;~
FG=Figma;~
VG=Vega;~
VL=Vega-Lite;~
OB=Observable;~
SB=Seaborn;~
GS=Google Suite;~
PPT=PowerPoint.
\end{minipage}
\vspace{-2.5em}
\end{table*}

We conducted a two-phase interview study. In Phase 1, we interviewed visualization practitioners to surface the tacit knowledge and annotation design considerations that guide their work. In Phase 2, we interviewed visualization educators and researchers to place practitioner accounts in dialogue with pedagogical and theoretical perspectives, surfacing boundary conditions and risks that practitioners may not articulate explicitly.

\subsection{Phase 1: Interviews with Visualization Practitioners}

\paragraph*{Participants.} We conducted semi-structured interviews with ten practitioners (P1--P10; \autoref{tab:participants}) who regularly create annotated visualizations. We recruited participants through social media, visualization-focused Slack communities, and the Data Visualization Society’s \href{https://www.datavisualizationsociety.org/research-recruitment}{DataViz Research Recruitment} program. 
We used purposive sampling~\cite{palinkas2015purposeful} to capture variation in practitioner roles, work contexts, tools, outputs, and intended audiences (\autoref{tab:participants}). We stopped recruitment when additional interviews reinforced existing considerations without introducing new ones.

\begin{table*}[!t]
\caption{Educator participants (E1--E7).}
\label{tab:educators}
\centering

\scriptsize
\setlength{\tabcolsep}{2.6pt}
\renewcommand{\arraystretch}{1.02}
\rowcolors{2}{gray!6}{white}

\resizebox{\textwidth}{!}{%
    \begin{tabular}{@{}l l c p{0.075\textwidth} p{0.1\textwidth} >{\raggedright\arraybackslash}p{0.35\textwidth}@{}}
\toprule
\rowcolor{gray!12}
\textbf{\textsc{ID}} &
\textbf{\textsc{Role}} &
\textbf{\textsc{Years teaching vis.}} &
\textbf{\textsc{Gender}} &
\textbf{\textsc{Region}} &
\textbf{\textsc{Annotation coverage in course}} \\
\midrule
E1 & Associate Professor & 10+   & Man   & Europe        & Integrated across topics \\
E2 & Associate Professor & 10+   & Man   & North America & Integrated across topics; reinforced through examples \\
E3 & Professor           & 10+   & Man   & Europe        & Integrated across topics; reinforced through examples \\
E4 & Associate Professor & 6--10 & Woman & North America & Mostly implicit; reinforced through examples \\
E5 & Associate Professor & 1--2  & Man   & North America & Brief coverage embedded in human-centered design/UX courses \\
E6 & Retired Professor   & 10+   & Man   & North America & Mostly implicit; reinforced through examples \\
E7 & Research Scientist  & 10+   & Man   & Europe        & Mostly implicit; reinforced through examples \\
\bottomrule
\end{tabular}%
}
\vspace{-3em}
\end{table*}

\paragraph*{Procedure.}
We conducted remote semi-structured interviews via videoconferencing, following a driver--navigator protocol~\cite{akbaba2023two} in which one author led the session while a second monitored the guide, took notes, and asked follow-up questions. Each session began with questions about the participant's role and workflow, then used artifact walkthroughs to ground the discussion in concrete examples~\cite{harper2002talking}. Participants typically screen shared and discussed 5--7 annotated artifacts (dashboards, static charts, interactive charts, and infographics) created for different media and audiences (e.g., print or web; policy stakeholders, executives, and general readers). Using these artifacts as prompts, we asked when and why participants add annotations, what they choose to annotate, and how placement, density, and styling vary with audience, medium, and tools; we also probed tool constraints, organizational standards, and how participants evaluate and revise annotations. In recruitment, we used a broad definition of \emph{annotation} to capture participants' own interpretations; at the start of each interview, we described annotations as additional textual or graphical elements added to a visualization and asked participants what they considered an annotation (Section~4.1). Interviews lasted 45--75 minutes (mean 60). Sessions were audio-recorded, transcribed, and reviewed for accuracy by the first author. After each interview, both interviewers wrote structured reflexive notes and held a brief debrief, which we retained as part of the audit trail used during analysis~\cite{nowell2017thematic}.

\subsection{Phase 2: Interviews with Visualization Educators}

\paragraph*{Participants.}
We conducted a second set of semi-structured interviews with seven visualization educators and researchers (E1--E7; \autoref{tab:educators}). We again used purposive sampling to recruit information-rich participants with substantial experience teaching visualization. We invited 50 educators via email through our professional network; seven agreed to participate. We designed Phase~2 as a structured peer debriefing~\cite{spall1998peer,guba1982epistemological}, in which educators critiqued the practitioner-derived considerations to surface boundary conditions, risks, and teaching-oriented framings. Given the focused aim of eliciting educator reflections on six prompts and the specificity of the expert sample, seven interviews provided high information power for this targeted inquiry~\cite{malterud2016sample}, consistent with evidence that key themes often emerge within the first several interviews in focused qualitative studies~\cite{guest2006many,baker2018many}. Participation was voluntary and uncompensated in both phases, and the protocol received institutional ethics approval.

\paragraph*{Procedure.}
Interviews were conducted remotely via videoconferencing and followed the same driver--navigator protocol as Phase~1. After brief background questions (role, teaching context, and whether annotations are addressed in their courses), we discussed the practitioner-derived considerations one by one (\autoref{fig:teaser})
using short Phase~1 summaries as prompts. For each consideration, we asked whether educators agreed, where it may break down, how they would frame it in teaching, and what risks or missing context they saw. The protocol also invited participants to connect specific considerations to broader pedagogical and design concerns. Sessions lasted 25--50 minutes (mean 35).

\vspace{-5pt}
\subsection{Data Analysis}
We analyzed anonymized transcripts from both phases using \textit{reflexive thematic analysis (RTA)}~\cite{braun2006using,braun2019reflecting,braun2021one}. RTA treats coding and theme development as interpretive, researcher-led work; accordingly, we do not report inter-rater reliability and instead support transparency through reflexive memoing, an evolving codebook, and traceable links between excerpts and themes. We report procedures using the COREQ 32-item checklist as a reporting standard~\cite{tong2007consolidated}. We also describe coder roles, theme development, and traceability materials (interview scripts, reflexive notes, codebook, and edited/unedited excerpts), consistent with interpretivist rigor guidance discussed in visualization research~\cite{meyer2019criteria}.

\textbf{Phase 1.} After each practitioner interview, the two interviewers wrote brief structured reflexive notes and held a short debrief to capture initial interpretations, questions, and notable examples. The first author then read all transcripts and applied semantic codes to segments describing annotation practices, constraints, and rationales. We maintained an evolving codebook with code definitions and representative excerpts. Codes were iteratively grouped into candidate themes and refined into the six considerations.

\textbf{Phase 2.} We used the practitioner-derived considerations as prompts in educator interviews and as sensitizing concepts during analysis. We coded educator transcripts to capture how educators endorsed, qualified, challenged, or extended each consideration, and to record boundary conditions and risks that shaped how we described the considerations in the paper.

\textbf{Team involvement and theme scope.} Across both phases, the first author led coding and theme development. The two interviewers conducted a structured peer debriefing, reviewing each consideration against the linked excerpts in the codebook and the reflexive notes. Disagreements were resolved by returning to the relevant transcript segments and refining code definitions or the wording and scope of the considerations. Some excerpts informed more than one consideration; in those cases, we recorded cross-links and specified how the excerpt supported each consideration, rather than forcing a single assignment.

We lightly edited quoted material for readability only: we corrected minor grammatical errors, removed disfluencies (e.g., ``um'', ``you know''), and excerpted longer passages while preserving meaning. \textbf{\href{https://osf.io/y5zw3/overview?view_only=91f4d45f6c1440138c4efac2ca011399}{Supplementary materials}} include
(i)~interview scripts;
(ii)~Phase--1 reflexive notes;
(iii)~32-item COREQ checklist;
(iv)~edited quotes paired with the corresponding unedited excerpts; and
(v)~the codebook.

\section{Findings}
\label{sec.results}

\subsection{What is Annotation to Practitioners?}
Practitioners described annotation as an additional layer on top of base data elements, separate from the main encodings but still part of the same visualization. Across all ten practitioners (P1--P10), elements such as text labels, callouts, arrows, highlights, enclosures, and, in some cases, titles, subtitles, and captions were treated as annotations when they changed how people read the chart. Practitioners sometimes referred to an \textit{annotation layer} and moved elements into or out of it, depending on whether they were making a claim, drawing attention to a specific aspect, or providing context. In this sense, an annotation is defined by its function rather than by a fixed set of visual forms. The same element type can be part of the base encoding in one chart and an annotation in another.

\vspace{-8pt}
\subsection{Why do Practitioners Annotate?}
Practitioners reported using annotations to emphasize key findings and supply essential context, to guide reading through a clear story, and to reduce misinterpretation when charts circulate without accompanying text.

\paragraph*{Emphasis and context.}
All ten practitioners reported using annotations to highlight the values, events, or regions that matter most and to provide sufficient context so that a visualization can be understood on its own. P6 estimated that \quoteP{a good, strong 80\% of the reason why I do it is to draw attention to things that need to be taken action on}. P5 used annotations to make scope and caveats visible when charts travel, placing notes so \quoteP{if you screenshotted the overall chart, you would see a comment} and framing this as a responsibility in a setting where \quoteP{everyone screenshots things} and charts can be reshared without their original context. P9 wanted the visualization to be \quoteP{complete within one image} and tried to \quoteP{make one image convey as much as possible ... enough information but also simple and minimal looking chart} so that a single view communicates both key values and essential background.

\paragraph*{Narrative guidance.}
A majority of practitioners (P1--P3, P5--P9) used annotations to set reading order and move from exploration to explanation. For example, P6 placed a brief framing sentence at the top so readers know what to look for \quoteP{a line or two up top to describe what they need to look at}. P3 preferred direct, proximal labeling, so the path through the figure is clear, and noted that hover interactions are unreliable on mobile. P8 tied the number and placement of callouts to the narrative, noting that \quoteP{it just has to support the logical flow … contribute to the understanding}, and kept phrases short to maintain prominence without crowding.

Some practitioners (P3, P5, P7, P9) also embedded brief caveats or scope notes to ensure figures remain interpretable when copied, cropped, or reshared.

\subsection{Annotation Design Considerations}
We present six annotation design considerations that capture recurring production decisions practitioners described when creating annotated charts (\autoref{fig:teaser}). The considerations emphasize how designers operationalize annotation intent within the constraints of audience, medium, and tool. We relate the considerations to prior work on visualization design and reader effects to situate them in the existing literature. Educator reflection serves as a structured peer debriefing~\cite{spall1998peer} to surface boundary conditions and trade-offs rather than to endorse a single best practice.

\vspace{-4pt}
\noindent\circnum{1} \textbf{Tailor annotations to audience knowledge and stakes.} Prior work shows that annotations and textual framing shape chart interpretation, affecting perceived message, trust, and recall~\cite{hullman2011visualization,kong2018frames,kong2019trust}, and that takeaways depend on how charts and text are integrated and on the amount of textual support provided~\cite{kim2021towards,stokes2022striking,stokes2023role,segel2010narrative}. These effects are documented largely in controlled settings; our findings instead describe how practitioners reason about these trade-offs under real production constraints, navigating a recurring decision: what to explain, what to foreground, and what context to make explicit on the chart, given the audience and the stakes of misinterpretation.

Practitioners began by identifying who would read the chart and what those readers needed to extract from it. Most (P1, P3--P10) tied wording, text quantity, and placement to audience knowledge, role, and goals. P4 started by asking \quoteP{Who is the audience? What data do you have?} and then adjusted annotation density and phrasing. P6 contrasted executives and peers: \quoteP{management just needs to look at the total numbers, and the biggest chunks to pay attention to}, whereas colleagues responsible for a process receive a chart that is \quoteP{more microscopic}. P1 designed for non-specialists, so annotations \quoteP{communicate to the ordinary person}. P7 and P8 added explanation for readers who \quoteP{might not be as familiar with some of the measures} and, in presentations, highlighted \quoteP{the stuff that is most interesting or relevant} rather than every detail. P9 aimed to have a single image \quoteP{convey as much as possible} while keeping annotation minimal enough that people do not feel overwhelmed. P10 tailored labels for clients who \quoteP{really want to see the full numbers} versus those who \quoteP{just want the story}, using different mixes of on-chart notes and tooltips.

When charts were likely to circulate or when misinterpretation would be costly, several practitioners (P3, P5, P7, P9) used annotations as a \emph{defensive layer}, embedding essential assumptions, caveats, or context on the chart so it remains interpretable when detached from the surrounding narrative. P3 linked this to the \quoteP{curse of knowledge}~\cite{xiong2019curse}, noting that what is obvious to authors may not be obvious to readers, and \quoteP{that's where we end up adding a bunch of annotations}. P5 designed explicitly for redistribution: \quoteP{everyone screenshots things and just shares them without the original source}, so she embedded comments so \quoteP{if you screenshotted the overall chart, you would see a comment}. Practitioners also constrained defensive notes to preserve scanability: P9 applied the same constraint, wanting a clean image that still carried essential context at a glance, and P8 required text to be \quoteP{short, easy to read, non-distracting, and super germane} to avoid introducing clutter. This pattern aligns with work on misleading communication, which shows that charts can circulate stripped of provenance and that designers should anticipate strategic reuse~\cite{fan2022annotating,lisnic2023misleading}.

Educators likewise treated audience tailoring as central, but emphasized limits and costs of defensive annotation. E3 questioned the need to \quoteE{add defensive labels all over the place}, warning that they can create an unsustainable maintenance burden and recommending clearer structure and targeted scaffolding instead. E6 endorsed embedding additional context when redistribution risk is high, but only when the information is essential to the chart’s main claim. E7 noted that while clarification can help, no design decision can completely prevent deliberate misuse.

\Takeaway{Start by specifying the audience and what they must extract from the chart. When redistribution risk is high, embedding essential caveats or context directly on the chart can help. Limit such notes to what is truly necessary, keep them short, and weigh the maintenance cost against the benefit.}

\vspace{-2pt}
\noindent\circnum{2} \textbf{Create hierarchy among multiple annotations.}
With multiple annotations, readers must decide what to attend to first and how annotations relate to the data. Prior work frames this phenomenon as a visual hierarchy and reading order problem, where authors use ordering and emphasis cues to guide attention and interpretation~\cite{Ware:2004:IVP,segel2010narrative,hullman2013deeper}. Empirical studies of communicative chart design suggest that clarity improves when designers focus attention on a small set of elements and reduce competing detail~\cite{ajani2021declutter}, and that overusing emphasis can reduce effectiveness by creating interference~\cite{strobelt2015guidelines}. Work on overlays and presentation cues similarly shows that layered cues can aid chart reading when applied selectively and consistently~\cite{kong2012graphical,kong2017internal}. Practitioners described hierarchy as the mechanism they use to control attention when many annotations are possible.

Practitioners treated hierarchy as essential, especially in dense charts (P1--P4, P7--P10), so viewers know what to read first without being overwhelmed by competing notes. They usually chose one or two primary statements and reduced the prominence of the rest. P10 made this explicit, noting that \quoteP{it is a hierarchy, and that affects where and how I place them,} and aimed for one main takeaway in the title, with secondary annotations that are \quoteP{not all screaming at you at the same time}. Practitioners encoded this priority through emphasis: P1 styled the key note to \quoteP{catch your eye the moment you look at it}. P7 used highlights and bold colored text to \quoteP{grab attention to the piece of info that is important}, keeping other notes lighter. P9 put \quoteP{a little box around the important text} and left other comments as plain text. Hierarchy was also enforced through revision: P3 and P8 populated charts, then checked \quoteP{if things are busy, if it’s clear}, adjusting label frequency, color, and size, and \quoteP{tak[ing] as much away as possible} while keeping the main takeaway prominent. P4 limited annotations, for example \quoteP{the three largest growth, and then the three largest shrinkage}, and warned that undisciplined labeling can \quoteP{take the natural flow of the reader's eye away from what's important}.

Educators reinforced the same goal of establishing a clear reading order anchored in the data, and emphasized how to make it checkable. E1 asked students to make \quoteE{the most important annotation stand out the most} while preserving a readable order. E2 advocated a gradient of emphasis where \quoteE{the secondary ones have a smaller visual emphasis}. E4 stressed that hierarchy should match the intended sequence of interpretation, asking whether \quoteE{people are visiting annotations in the orders that I expect them to}. E6 similarly emphasized how \quoteE{the order of those labels, horizontally or vertically} shapes understanding. Hierarchy, across both groups, is not about adding more notes; it is about coordinating salience and placement so attention moves through the chart as intended.

\Takeaway{Treat multiple annotations as an attention plan anchored in the data. Choose one primary annotation and give it the strongest placement and emphasis so it is read first. Keep remaining annotations consistently secondary through lighter styling and reduced salience. If any secondary annotation competes with the primary or disrupts the reading order, remove, merge, or restyle it.}

\vspace{1pt}
\noindent\circnum{3} \textbf{Place annotation text next to data before using connectors or legends.}
Placing annotation text is fundamentally a \emph{label-target association} problem: viewers must link words to specific marks or regions, and greater separation increases visual search and association effort~\cite{Ware:2004:IVP}. Empirical work shows that placement can shift takeaways, and that some information is better placed near the data than in a title~\cite{stokes2022striking}. Prior work maps the placement design space, including taxonomies of external labels and leader lines~\cite{bekos2019external} and readability trade-offs across connector styles~\cite{barth2019readability}. In our interviews, practitioners described a pragmatic fallback order: place text adjacent when possible; otherwise attach it with a short connector or small enclosure; use a keyed list or legend only when direct attachment would crowd the view or make the mapping unclear.

We use \textit{annotation text} to refer to the words or symbols that explain something, \textit{target elements} to refer to the marks or regions they describe, and \textit{connectors} to refer to lines or arrows that link text to targets. Nine practitioners (P1--P6, P8--P10) prioritized placing annotation text adjacent to target elements to reduce association effort. P8 was a \quoteP{big fan of not using legends if possible and directly annotating the features themselves}, noting that legends create unnecessary eye movement. P4 preferred to \quoteP{put the annotation where the action is, like directly on the dot [data element]}. Educators echoed this rationale. E4 observed that \quoteE{putting something spatially beside the data means people do not have to hold things in working memory, it just makes sense.} E2 cautioned that labeling every mark can backfire: \quoteE{if I try to put a label on everything, at some point I might as well just use a table.} In practice, practitioners reserved adjacent labels for the most important targets; when there were too many, they reduced marks, aggregated, or changed the view. Educators also highlighted boundary conditions. E6 suggested in situ labels for broader audiences or higher-stakes contexts. E7 noted that adjacency can be brittle in interactive or three-dimensional views with motion or occlusion~\cite{ponchio2020effective}.

When adjacency was infeasible due to space, practitioners (P1--P4, P6, P10) used short leader lines or arrows to preserve association without overlap. P8 reiterated a preference for direct labeling but noted that connectors become necessary when labels would collide. Educators added two refinements. First, E1 suggested lightly enclosing a small group of related elements and placing one annotation inside the enclosure, rather than drawing many separate connectors. Second, E3 emphasized connector styling for clarity: \quoteE{all my leader lines I would make them as visually distinct as possible}. These refinements align with evidence that connector style affects readability and label-target assignment performance~\cite{barth2019readability}. Across accounts, connectors and enclosures were kept short, uncrossed, and visually distinct from data marks; when connectors began to dominate the view, practitioners preferred reducing the number of annotated targets or switching to a keyed approach.

Nine practitioners (P1--P6, P8--P10) avoided detached legends or keyed lists. Some, such as P7 and P10, placed notes just below charts. P2 used mobile fallbacks with floating annotations or scroll in notes keyed by number, reflecting constraints in responsive visualization settings~\cite{hoffswell2020techniques,kim2022cicero}. P8 weighed the cost: \quoteP{a legend in a figure is a lot more cognitive burden than a well placed annotation}, using legends only when \quoteP{absolutely forced by the content.} Educators noted exceptions for distance and stability. As E1 said, \quoteE{if you have three things that share some property and they are far apart, a single legend entry might make sense.} E7 recommended legends in small multiples or interactive scenes. P3 suggested a scannability rule: \quoteP{orient the legend such that it matches the order of the data points.}

\Takeaway{Annotate the highest-priority targets next to the data. When proximity is not feasible, use short, uncrossed connectors or a single enclosure to preserve unambiguous mapping. Use a key or legend only when distance, interaction, or quantity makes attachment unstable, and order it to match the chart’s reading order.}

\noindent\circnum{4} \textbf{Use color and style to associate the label with the data element.}
Associating annotation text with its referent is a correspondence problem: color matching is reliable only when differences remain discriminable for the mark types in use and under realistic viewing conditions~\cite{healey1996choosing,szafir2017modeling}, and palette choices are further constrained by medium and reproduction~\cite{harrower2003colorbrewer}. Because correspondence can fail under these constraints, practitioners treated color, type weight, and small graphical cues as a deliberate association scheme, avoiding color-only matching and checking legibility and contrast, especially when designs must work across audiences and settings~\cite{elavsky2022accessible,joyner2022visualization}.

Color and style matching was a recurring strategy for linking annotations to data elements (P2--P4, P6--P8, P10). P2 described a planetary chart where the label color matched the data point exactly, noting that \quoteP{the Planet Nine is the same blue [as] it is, like, in the circle}, so readers could connect label and point without consulting a legend. P6 similarly noted, \quoteP{So, typically, so visually, I match the annotation to the color} of the bar while adjusting size so that people \quoteP{furthest in back} can still read it. P3 explained that their team \quoteP{tr[ies] to stick to brand guidelines in terms of colors} and sometimes \quoteP{tone that down a little bit for the color matching annotation to be more visible, more accessible} rather than introducing a competing palette. P8 reported copying exact RGB values so \quoteP{the color of the letters} matches the visual element, and pairing this with proximity and occasional pointer lines so \quoteP{what goes with what} is immediately clear. In each case, matching functioned less as decoration and more as a repeatable scheme: consistent styling signaled which elements belong to the annotation layer, while additional cues reduced ambiguity when color alone was insufficient.

Educators converged on the value of consistent styling and controlled redundancy, but differed on how tightly labels should match the data. E1 described an example where color, position, and connector ticks align, arguing that a \quoteE{super well designed annotated chart} should use cues redundantly. E5 similarly noted that \quoteE{the biggest benefit of annotation is that you can leverage redundancy}. Educators also emphasized constraints. E4 and E6 treated color matching as appropriate only when the palette remains small enough to stay discriminable. E2 and E3 preferred annotation styles that are coherent with one another but not necessarily identical to the data, reflecting that annotations can function as a distinct information layer. E7 cautioned against coloring text to match filled marks, recommending swatches or boxes for association and reserving text color mainly for emphasis. Both groups treated color as a redundant cue rather than a primary one, consistent with evidence that discriminability and contrast cannot be assumed across mark types and viewing conditions~\cite{szafir2017modeling,elavsky2022accessible}.

\Takeaway{Treat annotation--target association as a deliberate scheme. Use color matching only when categories remain clearly discriminable, and pair it with a non-color cue such as proximity, a connector tick or line, a swatch, or an enclosure. Keep text legible with sufficient contrast, and avoid color-matched text on filled marks when it reduces readability.}

\vspace{2pt}
\noindent\circnum{5} \textbf{Decide whether annotations blend with or stand apart from data.}
Whether annotation elements should read as part of the encoding or as author-supplied commentary is an explicit design decision, not a purely aesthetic one. Prior work shows that layered elements support chart reading when their visual role is clear and consistently applied~\cite{kong2012graphical,kong2009perceptual,kong2017internal}, but that highly salient treatments can interfere with other cues~\cite{strobelt2015guidelines}. In narrative settings, annotations are a primary mechanism for adding context and emphasis~\cite{ren2017chartaccent,hullman2013contextifier,bach2018design}, and framing work cautions that such additions can steer interpretation~\cite{hullman2011visualization}. Practitioners framed this as a layer-identity choice governing how readers parse each annotation element relative to the underlying data.

Practitioners described two contrasting strategies. In integrated designs, annotations share typography, color, and line style with the marks so the figure reads as a single object. P1 and P3 emphasized adhering to organizational brand guidelines, noting that \quoteP{so far as you are following the brand guide of the company, you are good to go} and that once fonts and colors are consistent, \quoteP{your annotation is good}. P2 described print graphics where annotation text uses the \quoteP{same color, same type style so it feels like one thing}. P9 sometimes muted annotation text but kept it in the \quoteP{same family as the rest of the design} so that \quoteP{an uninformed viewer might not even realize those labels were added later}. Other practitioners intentionally made annotations visually distinct to signal added context. P5 used bright red arrows and watermark-like notes to make it \quoteP{obvious those were not part of the original chart}. P7 added boxes or callout shapes so readers see the text as \quoteP{extra context}. P10 treated time-break and event markers on time plots as elements that \quoteP{sit on top as their own layer}.

Educators rejected any fixed rule and framed the choice in terms of interpretation, reuse, and maintenance. E5 asked: \quoteE{Do you want people to maybe think the annotation is part of the data or do you need it to be obviously an added layer?} E3 proposed a consistency test: \quoteE{things that have similar characteristics look the same, and things that have different characteristics look different,} advocating stable styling within a role and clear differentiation across roles. E2 similarly recommended giving annotations a stable visual identity across related figures so they remain distinguishable when charts circulate. Educators also emphasized costs and failure modes: E1 cautioned that deeply integrated styles can look polished but are expensive to maintain across revisions. E5 warned that highly salient annotation styles can overpower the data. E7 highlighted interactive and three-dimensional scenes, where occlusion and changing viewpoints blur layer boundaries, and argued for placing annotations in fixed panels or legend-like regions with distinct styling rather than attaching many moving labels directly to 3D data elements~\cite{ponchio2020effective}.

\Takeaway{Decide the role of each annotation before styling it. Use an integrated style when annotations should read as part of the chart; use a distinct style when they should read as added context, but keep it visually subordinate to the data. Reuse both styles consistently across related charts so layers are immediately recognizable and revisions remain manageable.}

\vspace{2pt}
\noindent\circnum{6} \textbf{Scale annotation count to chart clarity and available space.}
Annotation density trades off explanatory support against clutter. Empirical work on communicative charts supports focusing attention on a small set of focal elements while reducing competing detail~\cite{ajani2021declutter}, and studies of chart–text integration show that text amount and placement can shift reader takeaways~\cite{stokes2022striking}. Feasible density is also constrained by medium and interaction: dashboard studies report annotation constraints under limited screen space~\cite{elias2012annotating,badam2022integrating}, responsive visualization changes what remains legible as layouts shrink~\cite{hoffswell2020techniques,kim2022cicero}, and communicative authoring often requires cross-tool iteration~\cite{bigelow2016iterating}. Practitioners treated density as context-dependent, shaped by chart complexity, available space, medium, and revision effort, citing numeric rules of thumb as local guidance rather than universal limits.

Practitioners described two complementary workflows. In an \textit{additive} approach, they started with a clean chart and added annotations until the display felt busy. P1 aimed for \quoteP{on average, three, maximum five, if I'm pushed} and folded extra explanation into existing callouts. P3 used colleague feedback as a stopping cue, asking \quoteP{if something looks too busy or if we could have done fewer labels}. In a \textit{subtractive} approach, they began with more annotation and then pruned. P4 and P8 annotated many candidate events, then kept \quoteP{whatever number help[s] the logical flow of the story} (P4) and \quoteP{tak[ing] as much away as possible} (P8). Tooling shaped which workflow was practical: cross-tool pipelines favored incremental addition, while bespoke views built in libraries such as D3~\cite{bostock2011d3} made broad labeling followed by pruning easier.

Practitioners also adjusted density based on the medium and available space. P2 and P10 reduced text on small or responsive displays, where \quoteP{it actually just gets smaller and smaller} and teams \quoteP{rely more on concise labels and tooltips}, while reserving longer explanations for PDFs or static reports. P6 limited dashboards viewed on a single management screen to \quoteP{one or two annotations}. P7 and P8 used more annotations in slides or talks where brief callouts guide attention. When there was no room to label marks directly, P7 and P10 used short notes or keyed captions beneath narrow plotting areas, and P2 moved longer explanations into surrounding prose, noting \quoteP{we like nice visualization, less text} on small screens. Interactivity shifted what needed to be persistent: for interactive work, P4 emphasized relying on hover for details on demand~\cite{shneiderman2003eyes}, where \quoteP{it’s going to be the tooltip that tells you} additional details, keeping the persistent layer sparse to avoid clutter. For static artifacts expected to circulate independently (e.g., infographics, reports, PDFs), practitioners (e.g., P3, P4, P5) accepted higher visible density to keep the chart interpretable when detached from accompanying text, while still trimming labels that impeded scanning.

Educators cautioned against treating practitioners’ numeric rules as fixed heuristics. They emphasized that density should be driven by the chart, task, and medium. E1 noted \quoteE{there isn't really a good rule} because the appropriate number depends on the figure, story, data, and space. E2 argued that three annotations can be far too few in maps. E3 stressed that the goal is \quoteE{complexity communicated with clarity. And that might involve one annotation, and it might involve 100}. E4 and E5 treated numeric guidance as a rule of thumb at most, and E6 and E7 endorsed ``less is more'' as a useful instinct but emphasized that designers should understand the rule well enough to know when to break it.

\Takeaway{Decide an annotation budget for the chart’s intended viewing context. Keep only the annotations that must stay visible to deliver the main claim on first view, and push secondary detail to interaction or surrounding text. Either add annotations until the display feels busy, or start broad and prune, removing, merging, or demoting until scanning remains effortless.}

\subsection{Medium, Evaluation, Tooling, and Teaching Annotations}
\paragraph*{Annotations in Static versus Interactive Charts.}
Practitioners framed annotation as medium-dependent: interactive systems can distribute explanation across interaction and state~\cite{heer2012interactive,kong2012graphical,subramonyam2018smartcues}, whereas static artifacts must remain interpretable without interaction and often circulate without a surrounding narrative.

Five practitioners (P1, P5, P6, P9, P10) treated static charts as a single opportunity to surface essential context. P5 layered callouts, captions, and defensive notes because \quoteP{the chart needs to carry the context itself}, and P9 aimed to have \quoteP{one image convey as much as possible}. Several (P1, P6, P10) exported a base chart and finalized annotation placement in presentation or design tools to keep key points legible without hover or click, consistent with communicative graphics that circulate as standalone images~\cite{segel2010narrative,rahman2023exploring,bigelow2016iterating}. Higher annotation density was the accepted trade-off for robustness under screenshots, reposts, and slide reuse.

Several practitioners (P3, P4, P6--P10) treated interactivity as an annotation layer, consistent with evidence that overlays help chart reading when cues are unambiguous and consistently accessible~\cite{kong2012graphical,heer2012interactive}. P3 offloaded explanation to tooltips while keeping only critical labels persistent, P7 used on-demand notes to keep dashboards clean, and P10 withheld labels until hover. P8 sequenced annotations with animation to pace interpretation, consistent with evidence that order and staged reveal shape chart--text integration~\cite{hullman2013deeper,zhi2019linking,zong2022animated}. Offloading explanation to interaction increases implementation and maintenance demands, since annotations must remain data-bound under filtering, sorting, and layout reflow~\cite{hoffswell2020techniques,kim2022cicero}. As P10 put it, \quoteP{labels need to move with the data.} Because hover is not guaranteed, practitioners kept the core message visible, with P3 insisting key points stay \quoteP{visible… from the get go} and avoiding content \quoteP{buried in the interactive web component}.

Educators agreed that annotation choices should adapt to the medium and emphasized complementary failure modes. E4 argued that \quoteE{forcing labels on everything can actually impede the experience} and that dynamic cues can guide readers without clutter. E1 warned against placing critical context behind hover, stating \quoteE{If important context is only on hover, many users will never see it.} E5 contrasted static outputs that must anticipate screenshots and circulation with interactive views that can reveal details progressively but still need to surface the intended message at the right moment. E7 pointed to repeated structures and three-dimensional scenes where labels can \quoteE{move or occlude each other} and argued that in those situations \quoteE{the legend makes a lot more sense}, aligning with documented challenges in robust 3D annotation~\cite{ponchio2020effective}.

\paragraph*{Evaluation of Annotations.}
Practitioners described lightweight evaluation as rapid critique-and-revision, matching how prior studies characterize visualization production under practical constraints~\cite{bigelow2014reflections,alspaugh2018futzing,parsons2020design,parsons2021understanding}. Eight participants (P1--P4, P6--P9) judged annotations based on whether the intended takeaway was immediate and whether confusion was reduced. P1 asked colleagues \quoteP{Did the annotation really communicate well... Did it bring more clarity?} and iterated on wording and placement until a typical reader could state the main point unaided. P6 used stakeholder questions as the stopping rule, finishing once those questions could be answered from the annotated view. Six participants (P1--P3, P7--P9) relied on rapid editorial or peer review to calibrate against authors overestimating what readers infer without guidance~\cite{xiong2019curse}; P3 recalled critiques where charts were \quoteP{too busy}, and P2 described editorial passes that refined wording and placement. Four participants (P1, P4, P7, P9) used nonexpert proxies, with P1 applying \quoteP{If I get two yeses and one no... If I get three yeses, then I am good to go.} P8 also used an informal visual check to remove notes that did not change interpretation. In deadline-driven contexts such as newsrooms (P2, P7), external testing was rare, though internal review usually preceded publication.

Educators accepted informal checks as a baseline, but wanted explicit tests of whether a first-time viewer could recover the intended message and whether each annotation justifies its attention cost. Chart--text studies show that adding text, changing wording, or shifting text relative to marks can change takeaways~\cite{kim2021towards,stokes2022striking,zhi2019linking}. Educators framed evaluation as managing cue competition, since selective overlays can help reading~\cite{kong2012graphical} while overly salient highlighting can distract or interfere~\cite{strobelt2015guidelines}. They advised scaling rigor to stakes: critique and peer review for routine work, and reader studies or eye tracking when annotations could steer attention or interpretation in consequential settings~\cite{alam2016analyzing,bryan2020analyzing}. They also treated this as an ethics issue, since emphasis can bias interpretation even when encodings are correct~\cite{correll2019ethical}.

\paragraph*{Annotation tooling.}
Practitioners described annotation authoring as multi-tool work, consistent with evidence that communicative visualization production often requires cross-tool iteration to achieve precise layout and typographic control~\cite{bigelow2016iterating}. Nine participants (P1, P2, P3, P5--P10) built a base chart in one environment and finished annotations elsewhere. P1 \quoteP{push[es] it into PowerPoint}, P6 would \quoteP{take a snapshot of Tableau and annotate it in PowerPoint} when defaults are too rigid, and P2 used Datawrapper to \quoteP{get an overview} and finalized in Illustrator. P5 built in Flourish and refined in Figma to \quoteP{put the annotation where I want}. P8, despite working in R, still \quoteP{push[ed] it into PowerPoint}, and P9 exported Flourish charts to Figma to adjust layout and non-English text. Participants framed export-and-finish as necessary to control proximity, hierarchy, and typography beyond platform defaults.

Practitioners framed multi-tool annotation as a trade-off: exporting bypasses rigid dashboard annotation features (e.g., Tableau, Power BI) for precise placement and styling, but breaks data linkage and accumulates maintenance debt~\cite{elias2012annotating,badam2022integrating}. Four practitioners (P3, P4, P6, P10) cited constrained placement and styling: P4 said these tools \quoteP{do not handle annotations very well} and leave authors \quoteP{really boxed in}, and P3 reported that one \quoteP{cannot put annotations exactly where you want}. P6 described losing \quoteP{the cognitive correlation between the graph and the information} when she could not match annotation text colors to encoded data. P10 reported that dashboards cannot reliably place labels across screen sizes and still \quoteP{move with the data}, consistent with responsive, data-bound layout challenges~\cite{hoffswell2020techniques,kim2022cicero}. Effort and organizational rules further constrained choices: P5 weighed whether annotation precision was worth the overhead, P7 valued tools whose \quoteP{point and click nature} enabled rapid revisions, and P4 and P6 were locked into client-mandated or organizationally required platforms regardless of preference. When built-in features were insufficient, most (P1, P2, P5--P9) exported static charts to PowerPoint, Illustrator, or Figma for a final pass, gaining control but forcing re-export and re-annotation on updates.

In interactive outputs, some relied on tooltips and kept only essential labels persistent; others (P3, P4) scripted in R or D3 to keep annotations data bound~\cite{bostock2011d3}. P4 overlaid HTML on dashboards to inject dynamic notes, acknowledging that \quoteP{most of the time it does not work}, and reserved full custom implementations for cases where annotations were \quoteP{as part of the visualization itself… just as core as the chart}. This aligns with systems that treat annotations as target-linked constructs with specified behavior rather than post-hoc artwork~\cite{hullman2013contextifier,ren2017chartaccent,rahman2025annogram,chen2025chartmark}. Educators echoed these constraints: E1 noted that students struggle when annotation demands exceed the expressiveness of mainstream tools, especially for precise placement and stable hierarchy during revision. E7 emphasized interactive, three-dimensional views where data-following labels can drift or occlude, making robust dynamic label management across toolchains difficult~\cite{ponchio2020effective}.

\paragraph*{Teaching Annotations.}
Educators reported that annotation design is rarely taught explicitly in visualization courses, which often foreground encodings, perception, and interaction~\cite{munzner2014visualization,Ware:2004:IVP}. Visualization education work also notes uneven curricular coverage and limited shared teaching resources, which can leave communicative practices implicit~\cite{bach2023challenges}. None of the seven educators had a dedicated unit on annotations. E4 noted, \quoteE{we don't really talk much about annotations in the class}, adding that it is \quoteE{more of an implicit on-the-fly thing than it is an explicit instructional piece in the way that I teach this.} E6 similarly reflected, \quoteE{I didn't really talk about that much in my courses} and that \quoteE{I and others assumed too much. It's just like, Oh, you just do the right thing} without formal guidance. E2 framed annotation as \quoteE{it's a second-level sort of consideration, in the sense that} early assignments emphasize mapping before students are asked to reason about explanatory text. As a result, students can finish a course without explicit practice in what to annotate, how to phrase it, or how to balance emphasis with clarity.

Educators described teaching annotations through critique embedded in existing assignments that force students to justify audience, primary takeaway, and how annotations should emphasize it. E5 encouraged students to \quoteE{think about your audience... what you're presenting, who it’s for, and how much complexity you are trying to convey}, positioning annotation as \quoteE{just one piece of the toolkit that visualization designers have} alongside other design principles. E2 described a scaffolded progression where early work omits annotations, but later he \quoteE{assume[s] they're going to do it} and penalizes missing essentials. E3 embedded annotation into reflection by asking students to analyze, record findings, then \quoteE{use those notes to make claims and explain [them] in annotations in a single, static graphic}, forcing a translation from exploration to communication. Critiques, projects, and studio assignments were the primary sites for repeated practice, often grounded in curated examples from practitioner-oriented books and blogs.

Practitioner-derived considerations gave educators concrete language for instruction and critique. Educators readily linked the considerations to topics they already teach, finding them useful as a shared vocabulary for critique rather than as a prescriptive framework. They resisted treating the considerations as rules. E5 aimed to \quoteE{avoid any prescriptive or rigid, you know, models or guidelines}, preferring to \quoteE{give these tools to students} and have them \quoteE{play around with it} while justifying choices. E7 described an annotation upper bound as \quoteE{a rule that has to be broken, you know, and will be broken ultimately}, stressing that designers \quoteE{need to know the rules to know when to break them}. This aligns with visualization literacy work that treats communicative competence as something developed through structured practice and feedback rather than absorbed as abstract rules~\cite{lee2016vlat,cui2023adaptive,ge2023calvi}.

\section{Discussion}
\label{sec.discussion}
\paragraph*{Interconnected Nature of Design Considerations.} Our findings suggest that annotation design considerations rarely act independently: changes in audience, medium, or communicative goal ripple through multiple choices, consistent with long-standing observations in visualization design~\cite{munzner2014visualization,parsons2021understanding}. Practitioners treated the considerations as a coupled set for a given chart, audience, and medium; resolving one decision constrained the others in ways that were difficult to capture through guidelines presented as separable principles. For example, tailoring text to audience knowledge and stakes shaped both the quantity of annotations and the depth of explanations. When they expected redistribution or higher-stakes use, they accepted denser and more explicit annotations, then relied on hierarchy and styling to preserve scanability, consistent with known trade-offs for overlays and highlighting~\cite{kong2012graphical,strobelt2015guidelines,stokes2022striking}. Placement, hierarchy, and styling were similarly intertwined. Positioning annotations near targets or in panels constrained how many notes fit and shaped reading order, while hierarchy depended on typography, enclosures, connectors, and emphasis working together, so a few anchor annotations carried the message~\cite{kim2021towards,zhi2019linking,hullman2013deeper}. Educators underscored that clutter, framing, and maintenance costs depend on how these choices are jointly resolved, especially during revisions and responsive layouts~\cite{bigelow2016iterating,hoffswell2020techniques,kim2022cicero}.

\paragraph*{Practitioner and Educator Perspectives.}
Practitioners and educators describe annotation design from related but distinct standpoints that mirror how visualization work is produced and taught. Practice-grounded studies show that design decisions are negotiated under deadlines, tool constraints, and iterative critique rather than applied as fixed prescriptions~\cite{alspaugh2018futzing,bigelow2014reflections,parsons2021understanding,parsons2020design}. Consistent with this, practitioners emphasized concrete constraints such as known audiences, specific platforms, and accountability when a chart misleads, adopting local, risk-driven strategies: they kept only a few visible callouts, used informal ``busy tests'' to manage density, placed text near targets before relying on connectors or keys, added context when charts may circulate as screenshots, and accepted denser overlays or brittle toolchains when stakes were high. Educators foregrounded legibility over time, maintenance cost, and the risk that heavy annotation narrows interpretation. Because they teach across domains and tools, they resisted turning practitioner patterns into rules, questioning numeric limits, rigid placement orders, and strong framing as general prescriptions. In their view, density, placement, and styling depend on the data, task, medium, and audience.

This juxtaposition separates situated tactics from their boundary conditions: practitioner accounts surface recurring production decision points, while educator reflections foreground failure modes and sustainability costs, consistent with structured peer debriefing as a rigor practice~\cite{spall1998peer}. This framing aligns with accounts of tacit and situated design practice, where judgment is context-dependent and iteratively refined under real-world constraints~\cite {polanyi2009tacit,schon2017reflective,suchman1987plans}.

\vspace{-7pt}
\paragraph*{Implications.}
The annotation design considerations in this paper are not a fixed checklist. We envision them as a shared language for critique, tool building, teaching, and future research, naming the decisions and tensions that experienced designers already navigate rather than prescribing resolutions. For tool builders, our results point to specific gaps between software and practice. Practitioners routinely export charts into PowerPoint, Illustrator, or Figma, losing data linkage and redoing annotation work when data or layout changes, suggesting that many tools still treat annotations as static text boxes overlaid on a completed chart. Our findings instead support representing annotations as first-class objects linked to target elements, with explicit control over placement and style. For educators, the interviews show that annotation is often handled implicitly or folded into other topics. The design considerations provide a framework for discussing annotation explicitly in existing courses, connecting it to audience, hierarchy, and medium, and asking students to justify their annotations and the strength of their framing, without turning these ideas into rigid rules. For researchers, the study offers a structured perspective on annotation design and its trade-offs, which can inform future empirical and systems work.

\vspace{-5pt}
\paragraph*{Limitations and scope.}
Our findings are based on interviews about self-reported practices and participant-shared artifacts, not on in situ observation of live authoring. This design supports reflective accounts of decision-making, but may omit aspects of practice that are difficult to recall or articulate. We also did not evaluate audience outcomes; the results summarize recurring reasoning and patterns in practice rather than causal effects on comprehension, persuasion, or trust, and should be interpreted as design hypotheses for future reader studies, field deployments, and workflow studies. Our sample covers diverse roles and sectors, but is not comprehensive. The ten practitioners and seven educators are experienced, and most examples concern two-dimensional charts and dashboards produced by individuals or small teams; collaborative analysis, three-dimensional visualization, and real-time monitoring were mentioned but not systematically sampled. Accordingly, the considerations are intended as transferable patterns for similar contexts, not universal norms, and may change across domains, organizations, visualization genres, and levels of expertise.

\section{Conclusion}
Annotation design involves consequential decisions that are rarely made explicit. Through a two-phase interview study with practitioners and educators, we surface these decisions as coupled choices shaped by audience, intent, medium, and tooling. Our practice-grounded considerations, paired with educator reflections, reveal the decision logic and boundary conditions that annotation taxonomies and isolated audience-effect studies do not capture. We offer these considerations as a shared vocabulary for design critique, tool and grammar development, teaching, and empirical research on annotation as an integral part of visualization design. Future work should validate these trade-offs through in situ authoring and reader studies, and translate the considerations into tools and teaching materials that support annotation design across contexts.

\setstretch{0.97}
{\footnotesize  %
\setlength{\itemsep}{0pt}
\setlength{\parsep}{0pt}
\setlength{\partopsep}{0pt}
\bibliographystyle{eg-alpha-doi}
\bibliography{_bib_abbrev}
}

\end{document}